\documentclass[12pt]{article}

\usepackage{amsmath}
\usepackage{amssymb}

\begin{document}

\title{Quantum Cryptography II: \\
\hspace*{-15.22843pt}\mbox{How to re-use a one-time pad safely even if P=NP}}

\author{Charles H. Bennett \\ \small(IBM Yorktown)\,\footnote{\,2014 update on email: \texttt{chdbennett@gmail.com}.}
\and Gilles Brassard \\ \small(Univ.\ de Montreal)\,\footnote{\,2014 update on email: \texttt{brassard@iro.umontreal.ca};
\hspace{\fill}\mbox{}
\mbox{\hspace{4em}\hspace{0.5mm}\,update on affiliation: also Senior Fellow of ETH-ITS and of CIFAR.}}\\
\and Seth Breidbart \\ \small(Box 1526, NY 10268)}

\date{November 1982 \\[-5ex] \mbox{}}
\maketitle

\begin{abstract}
When elementary quantum systems, such as polarized photons, are used to transmit digital information, the uncertainty principle gives rise to novel cryptographic phenomena unachievable %TYPO: was "unachieveable" in the original BBB82
with tradi\-tional transmission media, e.g.~a~communications channel on which it is impos\-sible in principle to eavesdrop without a high probability of being \mbox{detected}. With such a channel, a one-time pad can safely be reused many times as long as no eavesdrop is detected, and, planning ahead, part of the capacity of these uncompromised transmissions can be used to send fresh random bits with which to replace the one-time pad when an eavesdrop finally is detected. Unlike other schemes for stretching a one-time pad, this scheme does not depend on complexity-theoretic assumptions such as the difficulty of factoring. 

\end{abstract}

\vfill

\noindent
{\footnotesize
\textbf{Note written on 29 June 2014:}
This paper was written in November 1982 and
originally submitted to the \emph{Fifteenth Annual ACM Symposium on Theory of Computing}, but it was rejected.
Shortly thereafter, two of the authors (Charles H.\ Bennett and Gilles Brassard) discovered
what became known as BB84, which seemed like such a better idea that they gave up
on resubmitting this earlier work.
On~the occasion of the 30th anniversary of the BB84 paper, \emph{Natural Computing} has invited us
to get this paper finally published.
This freshly typeset version is scrupulously faithful to the original 1982 submission,
except for the correction of about one dozen typographical mistakes and a few footnotes written in retrospect by the authors.}

\pagebreak

\sloppy
\addtolength{\parskip}{0.5ex}

\section{Introduction}

In conventional information theory and cryptography, it is taken for granted that a digital message can always be copied easily, even by someone ignorant of its meaning. Analog messages (e.g.~handwritten signatures) are somewhat harder to copy, but not really infeasible, and digital data can be protected to a considerable extent by interposing a restrictive hardware interface between the data and the outside world (e.g.~smart credit cards); but in both these cases, the difficulty of copying is only technological, not fundamental. However, when elementary quantum systems such as polarized photons are used as the transmission medium, routine copying of messages is no longer possible even in principle. In particular, there are ways of encoding messages so that they can be copied reliably only with the help of certain key information used in forming the message. 

Quantum coding was first described in~\cite{W}, along with two applications: making money that is in principle impossible to counterfeit, and multiplexing two or three messages in such a way that only one can be read. More recently~\cite{BBBW}, quantum coding has been used in conjunction with public key cryptographic techniques to yield several schemes for unforgeable subway tokens. Here we show that quantum coding considerably enhances the usefulness of another standard cryptographic device, the one-time pad. 

Mathematically, a polarized photon acts like a two-bit read-once memory one of whose bits ($k$) serves as a read key for the other ($m$). Querying the memory with the correct $k$ yields the correct value of $m$. Querying with the wrong $k$ yields a random bit instead of $m$, and in either case querying resets the memory so that subsequent queries yield no new information. Even after a query, it is generally impossible to infer the initial state of \mbox{either} bit, because the memory gives no indication of whether its response was the correct response to the correct key or a random response to the wrong key. Because it represents the behavior %TYPO: was indeed "behavior" in the original BBB82, but was not corrected to "behaviour" in the final 2014 typeset version!  :-)
of an elementary quantum system, this kind of restricted-access memory should be thought of as a natural information-processing primitive, %TYPO: was "primative" in the original BBB82
not as a complex technological device that could probably be circumvented in principle. 

Ordinarily, when one thinks of a technological restricted-access memory, one has in mind an information-storage device. Photons can also be stored (e.g.~between mirrors, or in a closed optical fiber), but they cannot in practice be stored for very long, and their natural application is in the transmission of information. We thus have a situation in which restricted-access memory, as a storage device, is possible in practice but not in principle via conventional technology, and in principle but not in practice via storage of polarized photons. On the other hand, restricted-access transmissions, which can be read or copied only with the help of a key, are possible both in principle and in practice using polarized photons. 

\section{Essential Properties of Polarized Photons}

Polarized light can be produced by sending ordinary light through a polariz\-ing apparatus such as a Polaroid filter or Nicol prism. A~beam of polarized light is characterized by its polar\-i\-zation axis, which is determined by the orientation of the polarizing apparatus in which the beam originates. \mbox{Although} polarization is a continuous variable, and in principle can be measured as accu\-rately as desired by passing the polarized beam through a second polar\-izing apparatus, the uncertainty principle forbids measurements on any \mbox{single} photon from revealing more than one bit about the beam's polarization. In~particular, if a beam with polar\-i\-zation axis $\alpha$ is sent into a polarizer oriented at angle $\beta$, the individual photons behave dichotomously and probabilistically, being transmitted with probability $\cos^2(\alpha - \beta)$ and \mbox{absorbed} with the complementary probability $\sin^2 (\alpha - \beta)$. The photons behave deter\-min\-is\-ti\-cally only when the two axes are parallel (certain transmission) or perpendicular (certain absorption). %TYPO: was "absorbtion" in the original BBB82

If the two axes are not perpendicular, so that some photons are transmitted, one might hope to learn additional information about $\alpha$ by measuring the transmitted photons again with a polarizer oriented at some third angle; but this is to no avail, because the transmitted photons, in passing through the $\beta$ polarizer, emerge with exactly $\beta$ polarization, having lost all memory of their previous polarization $\alpha$. Any other elementary two-state quantum system, such as a spin-$1/2$ atom, behaves similarly dichotomously and probabilistically.

Another way one might hope to learn more than one bit from a single photon would be not to measure it directly, but rather somehow amplify it into a clone of identically polarized photons, then perform measurements on these; but this hope is also vain, because such cloning can be shown to be inconsistent with the foundations of quantum mechanics~\cite{WZ}.

\section{Quantum Coding}

In order to encode a message bit $m$ into a photon that can be read or copied reliably only with the help of a key bit $k$, we generate a photon with a selected one of the four polarization directions $0$, $45$, $90$ and $135$ degrees. [Generating a single photon of known polarization is possible by variation of the Einstein-Podolsky-Rosen setup~\cite{Bo}, in which a decaying atom emits two oppositely polarized photons. By polarizing and counting one photon, the other's presence is assured and its polarization fixed without measuring it directly.] If the key bit is a $0$, then the photon is polarized rectilinearly, i.e.~$0$ or $90$ degrees according to whether the message bit is $0$ or $1$. If the key bit is a $1$, then the photon is polarized diagonally, i.e.~$45$ or $135$ degrees according to the message bit.

\vspace{1.5ex}
\noindent
\emph{Def.}\ The \emph{quantum encoding} $Q_K (M)$ of a message $M$ by a key $K$ of equal length is the train of photons obtained by applying the above procedure bitwise to $M$ and $K$.
\vspace{1.5ex}

To read a quantum-encoded message with the help of its key, one simply reads each photon with a polarizer oriented so as to cause it to behave deter\-mi\-nis\-ti\-cally, for example, reading the rectilinear photons with a $0$-degree polarizer and the diagonal photons with a $45$-degree polarizer. An attempt to read a photon with the wrong key causes it to behave randomly, losing its stored information. For example, if a $45$-or $135$-degree photon is read with a $0$-degree polarizer, it will be transmitted with $50$ per cent probability in either case, and all evidence of its original polarization will be lost.

Suppose an eavesdropper intercepts and attempts to read a quantum transmission $Q_K(M)$ without being detected. Consider first the case in which the message $M$ and key $K$ are both random. Not knowing $K$, the eavesdropper makes the wrong measurement on half the photons, and thus obtains a message $M'$ differing from $M$ in $1/4$ of its bit positions (of course the eavesdropper does not know which ones). Having destroyed the original transmission by reading it, the eavesdropper must now, in order to remain undetected, inject a forged transmission designed to approximate the intercepted one as well as possible. Not knowing which measurements are wrong, the eavesdropper's best strategy is to produce a new train of photons in agreement with the results of the measurements, as if they had all been right. Half of the photons in such a forged transmission will be correct; the other half have wrong key values (i.e.~will be diagonal when they should be rectilinear, or vice versa), and when subsequently measured with the correct key by the intended receiver, these will give wrong answers half the time. Thus the error probability is $1/4$ per bit, both for reading the quantum transmission without knowing its key, and for having a forged replacement agree with what the original message would have said when decoded by the intended receiver. Of course, if the intended receiver knew only $K$ but had no prior knowledge of $M$, the eavesdropping would still\,\footnote{\,This word (``still'') appears to be superfluous. The authors do not understand in 2014 what they could have meant by it when they wrote the original 1982 manuscript.} go undetected, since a random message with random errors still looks random. Quantum money~\cite{W} corresponds to the case where the intended receiver (the bank) has perfect knowledge of both $M$ and $K$, while the counterfeiter knows neither. The usual message $M$ sent over communication channels is intermediate between these extremes: the receiver has partial prior knowledge of it (e.g.~expecting it to be in English).

Simply encoding an arbitrary message $M$ with a random quantum key $K$ has two disadvantages: 1)~if~the message is too random the receiver won't be able to detect eavesdropping, for the reason mentioned above; 2)~if~the message is too redundant (e.g.~English), eavesdropping will be detected, but by then the eavesdropper will have gained significant information about the message, perhaps even enough to decrypt it uniquely, because eavesdropping \mbox{induces} errors in only $1/4$ of the bits. (In this respect quantum coding differs from %TYPO: the word "from" was duplicated in the original BBB82
ordinary one-time pad encryption, where ignorance of the key prevents the eavesdropper from learning anything about the encrypted message,\footnote{\,In~2014, the authors realize that the phrase ``the encrypted message'' was ambiguous and confusing.   They intended it to mean the ``the message whose meaning had been concealed by encryption''---i.e.~the plaintext---rather than what would nowadays be seen as its more likely meaning in a cryptologic context, ``the message in encrypted form''---i.e.~the ciphertext.  Eavesdropping on a classical one-time pad transmission of course yields complete information on the ciphertext but none on the plaintext.} though of course it can be freely copied.)

We now define a stronger kind of coding that overcomes both these disadvantages. The trick is to make the message redundant with an error-detecting code $M \rightarrow E(M)$, then hide the redundancy from the eavesdropper by an ordinary one-time pad $J$, before applying quantum coding.

\pagebreak

\vspace{1.5ex}
\noindent
\emph{Def.}\ For any error-detecting code $E$ (assumed known to the eavesdropper) let the \emph{strong} \emph{quantum code} $S^E$ be defined as follows: let $J$ and $K$ be two %TYPO: the word "key" appeared here in the original BBB82
random key strings of length $|E(M)|$ not known to the eavesdropper.\footnote{\,In 2014, the authors noticed a possible ambiguity in this sentence. It~is the random key strings $J$ and $K$ that are unknown to the eavesdropper, not their length~$|E(M)|$.} Then the \emph{strong quantum encoding} $S^E_{J,K}(M)$ of message $M$ is the train of photons $Q_k(J \text{ xor } E(M))$.
\vspace{1.5ex}

It is obvious (because of the one-time pad $J$\,) %TYPO: was "one time pad" (without hyphen) in the original BBB82
that the eavesdropper can learn \mbox{nothing} about $M$ from $S^E_{J,K}(M)$. Moreover, for suitable error-correcting codes,\footnote{\,In 2014, the authors noticed that they had meant ``error-detecting codes'' here.} eavesdropping \mbox{incurs} a high risk of being detected. Even the rudimentary code of repeating the message twice $E(M)=MM$ suffices to detect eavesdropping with probability at least $1-0.79^k$ when $k$ photons have been intercepted, quite close to the optimum $1-0.75^k$ implied  by the independent, probabilistic nature of eavesdropping-induced errors.

Although the simple code $E(M)=MM$ is nearly optimal for eaves\-drop\-ping-detection, a more complex code would be preferable for another reason: the detection of deliberate message alteration. Although randomly quantum-coded photons cannot be read reliably, they can be altered reliably. For example, the polarization axis of a photon can be rotated by $90$ degrees, without measuring or otherwise disturbing it, by passing the photon through an appropriate sequence of mirrors (or, more mysteriously, through a sugar solution). If this manipulation were applied to the first and \mbox{$(n+1)$}st %TYPO there was an apostrophe before � st � here but not in the other place where $(n+1)$st  appeared; apostrophe removed here for consistency.
photons of a $2n$-photon transmission coded as above, both would be altered with certainty in such a way as to induce an undetected alteration in the message. A~more complex error-detecting code, e.g.~concatenating $MM$ with a check sum of the addresses of the ones in $M$, would make such alterations unlikely to escape detection. In the next section, where quantum transmissions are used to carry key information for future transmissions, it will be necessary to use an error-correcting code\,\footnote{\,In 2014, the authors noticed that they had meant ``error-detecting code'' here as well.} that provides  some `diffusion', in the sense of making each bit of $E(M)$ depend on many bits of~$M$\@. This prevents the eavesdropper who has luckily guessed a few bits of the present key from thereby efficiently inferring any bits of future keys. Finally, in section~\ref{practical}, we will need a code $E$ that corrects errors as well as detecting them, to make up for photons that arrive at the receiver but fail to be detected. 

\section{\mbox{Reusing a One-Time Pad Safely with the} \mbox{Help of Quantum Coding}}

We consider a situation in which two users of an insecure communications channel, who initially share a finite secret key, wish to communicate \mbox{secretly} as long as they can. In a classical channel, where eavesdropping is unde\-tect\-able in principle, they must assume that all their communications are being listened to, and the volume of safe communication is only \mbox{linear} in the size of the key, unless they resort to pseudorandom key-expansion %TYPO: was "key-expansions" in the original BBB82
schemes~\cite{BM,Y}, %TYPO: this comma was missing in the original BBB82
which are at best (assuming $\text{P} \neq \text{NP}$) only computationally secure. 

We show that by strongly quantum-coding their messages with suitable error-detecting codes, the sender and receiver can safely reuse the same $J$ and $K$ keys indefinitely until an eavesdrop is detected. (The safety is not absolute. There is an exponentially small chance ($O(2^{-|K|})$) that the eavesdropper, having guessed the entire quantum key $K$ correctly, will be able to eavesdrop on all the transmissions without detection and go on to break the reused $J$ key in the usual manner, as well as a moderate chance for the eavesdropper to learn %TYPO: the word "learn" was missing in the original BBB82
a few bits of the $K$ key correctly and go on to intercept and decrypt a few bits of each message; but these risks do not increase with the number of times an apparently secure key is reused.) An eavesdropper who tampers with or suppresses messages will also be detected with high probability, as will one who injects false messages. 

When an eavesdrop is detected, the sender and receiver can go on communicating with only slightly diminished safety by replacing their compromised keys by fresh random information sent over the channel in previous uncompromised transmissions. With high probability they will be able to continue communicating in this fashion for an exponential ($2^{O(|K|)}$) number of key changes,\footnote{\,Note added in 2014: this 1982 use of the asymptotic notation ``$O$'' was an example of the common physics usage, where it may mean either an upper or lower bound depending on context.
Here, we intended a lower bound, which in modern computer science usage would be denoted~$2^{\Omega(|K|)}$.} unless the eavesdrops become so frequent before then that they are forced to use up key information faster than they can replace it, in which case they will (with high probability) be able to cease communication before any of their transmissions have become uniquely decodable by the eavesdropper.

Because the replacement keys are truly random, rather than being pseudorandomly computed from an original seed, the security of the scheme would not be reduced by allowing the eavesdropper unlimited computing power. Neither would it be compromised by technological improvements in the art of eavesdropping. The scheme does incorporate one technologically unrealistic assumption, viz.\ that photons can be detected with perfect efficiency %TYPO: was "efficieny" in the original BBB82
(cf.~section~\ref{practical}, where this assumption is not made).

We sketch how these advantages can be achieved. The ability to send many messages without loss of security (when no eavesdropping is detected) follows from the exponential decline of the probability of escaping detection with the number of distinct bit positions ever subjected to eavesdropping, whether these bit positions are listened %TYPO: was "listend" in the original BBB82
to all at once, or a few at a time over the course of many transmissions. For this reason, a quantum channel could even be used to safely send arbitrarily many copies of the \emph{same} strongly coded transmission, without the eavesdropper being able to forge it accurately, provided the copies were sent one at a time, each only on confirmation that the preceding one had apparently not been listened to. By contrast, if many identical transmissions were sent all at once, the eavesdropper could intercept them all, reliably determine each polarization by multiple measurements, and then escape detection by forging many trains of photons with the now known sequence of polarizations. 

In order to be sure that no key is reused after a detected eavesdropping, the two communicating parties must alternate strictly in their use of each key. Otherwise the eavesdropper could, for example, intercept and absorb a message from A without  forwarding it to B and then wait for B to use the same key on a subsequent message. The effect of absorbing a message is thus the same as that of spoiling it through eavesdropping: neither party reuses the key with which it was sent. If the initial body of shared key information included several keys reserved for first use by A and several for first use by B, the parties could alternate in the use of each key without strictly alternating transmissions. Of course if multiple keys were in use, and particularly if some transmissions were being absorbed by the eavesdropper, the communicating parties would have to prefix each quantum transmission by a (cleartext) indication of which key it was encoded with, to avoid reading a message with the wrong quantum key and spoiling it.

The ability to change keys without serious loss of security depends on using a somewhat diffusive error-detecting code when new key information is transmitted. With the simple non-diffusive code $E(M)=MM$, an eavesdropper who by good luck has correctly guessed the first and \mbox{$(n+1)$}st bits of the current $J$ and $K$ keys %TYPO: was "key" in the original BBB82
will know what measurements to make to reliably read and forge the corresponding bits of a fresh pair of random keys $J'$, $K'$ when these are sent through the channel in four transmissions strongly encoded by $J$ and $K$; as well as confirming, by the consistency of decoding of the error-detecting code, that the guessed bits of $J$ 
and $K$ are indeed correct. Subsequent lucky guessing on further generations of keys (along with unlucky guessing causing some keys to be rejected due to detected eavesdropping) could be used to discover additional bits until, in linear time, some pair of keys $J''$ and $K''$ became entirely known.

To delay this collapse for an expected exponential number\,\footnote{\,Note added in 2014: as in the previous footnote, modern computer science usage would have us write an expected $2^{\Omega(n)}$ number of key changes.} of key changes $2^{O(n)}$, it suffices to use an error-detecting code that diffuses information about each bit of its argument $M$ among many bits of its value $E(M)$; so that knowledge, say, of any $n/4$ bits of $E(M)$ reveals little or nothing about any bit of $M$. Many error-detecting codes have this property, e.g.~a~random mapping from $n$-bit strings to $2n$-bit strings, or the linear code obtained by mod-$2$ multiplying $MM$ by an appropriate nonsingular matrix. With a diffusive code, knowledge of a few bits of $J$ and $K$ would not enable the eavesdropper to make reliable measurements of any bits of the replacement keys $J'$ and $K'$.

\section{A Practical Implementation}\label{practical}

Although visible light photons can be polarized with nearly perfect efficiency (e.g.~a~Nicol prism can split a beam into two beams, very nearly perfectly polar\-ized at right angles to each other, whose total intensity is scarcely less than that of the incoming beam), and transmitted with nearly perfect effi\-ciency (in a vacuum the only significant losses are due to diffraction, and these can be made negligible by using a beam diameter considerably greater than the square root of the product of the transmission distance and the wavelength of light), current technology allows them to be detected with only about thirty per cent efficiency.\footnote{\,Note added in 2014: this was the approximate quantum efficiency of photomultiplier tubes available
in~1982.}

Fortunately, the scheme of the preceding section can be modified to accom\-mo\-date finite detector efficiency, at the cost of using a more complicated error-correcting code $M \rightarrow E(M)$ in place of the error-detecting code, and a more complicated criterion for key rejection than the detection of a single error on decoding $E(M)$. Somewhat surprisingly, %TYPO: was "surpisingly" in the original BBB82
the modified scheme remains secure against an eavesdropper with a more efficient, or even perfectly efficient, photon detector. The volume of safe communication for this scheme is more than linear, but may be less than exponential, in the initial key size.

The main modification is to use standard faint pulses of polarized light instead of single photons, each pulse being of such a size that when it is sent into a detector of the given efficiency (e.g.~$30$ per cent), or split into several fainter pulses (e.g.~by~a half-silvered mirror, or a Nicol prism) and sent into several such detectors, the total number of photons detected obeys a Poisson %TYPO: was "Poissson" in the original BBB82
distribution of mean~$1$. Such a standard faint pulse can easily be produced by filtering a standard bright pulse of polarized light to reduce its intensity by the requisite constant factor.

A standard faint pulse of a given polarization resists copying nearly as well as a single photon would. The best strategy for an eavesdropper to copy a faint pulse is to use a half silvered mirror and two Nicol prisms to split the incoming pulse into four beams, one of each canonical polarization, and monitor each beam by a photon detector. Most of the time, only one of the detectors will register a photon, and the eavesdropper will be no better off than in the single photon case. Occasionally two or three detectors will register, yielding more information. Only when three detectors register will the pulse's polarization be known unambiguously (e.g.~if~both diagonal detectors and the vertical detector register, then the pulse must have been vertically polarized). The faint pulse works well because the chance of three detectors responding to the same pulse is only about $2$ per cent (for a Poisson distribution of mean $1$). The other $98$ per cent of the time, the eavesdropper does not learn the pulse's polarization unambiguously, and, as with single photons, cannot reliably copy~it. Even a technologically advanced eavesdropper, with perfectly efficient photon detectors, could not copy faint pulses reliably. For example, if the advanced eavesdropper uses $100$ per cent detectors to analyze %TYPO: was indeed "analyze" in the original BBB82, but was not corrected to "analyse" in the final 2014 typeset version!  :-)
a pulse intended for $30$ per cent detectors, an average of $3.3$ photons will be detected per pulse, but the chance that these will appear in three different beams, and thus reveal the pulse's polarization unambiguously, would still be only about $25$ per cent.

The converse phenomenon, namely statistical failure to detect even one photon when a pulse arrives, requires that the rejection test be made more complicated. Even if a transmission %TYPO: was "transmissions" in the original BBB82
is not subjected to eavesdropping, about $1/e$ of its light pulses go undetected, due to normal bad luck at the detectors. The rejection test must begin by deciding whether the number of missing light pulses is so great as to raise the suspicion of eavesdropping (a~wise eavesdropper now might not bother to forge replacements for the intercepted pulses, but instead let them remain missing, hoping to pass them off as pulses that arrived but were not detected). If the number of missing pulses is not too great, the error-correcting code must reliably restore the data they would have carried, as well as checking for polarization \mbox{errors}, which as before would indicate interception and forgery of some of the pulses. A~convolutional code~\cite{G} appears most suitable for achieving the desired high efficiency of error-correction in a channel with a large number of erasures~($1/e$). \mbox{Depending} on the purity of polarization available from the Nicol prisms, the code could be made to tolerate and correct a small number of polarization errors, but reject a larger number as evidence of forgery. Since the capacity of a binary channel with $1/e$ erasure probability is $0.632$, a four-fold expansion in $E(M)$ offers ample room for efficient error detection and correction. This in turn means that eight transmissions, each containing $n$ fresh key bits, would have to be accepted to replace the $8n$ bits sacrificed in a rejection.

The most problematical aspect of the modified scheme is the decision of when to reject a transmission and change keys. By contrast with the scheme of the previous section, it is now necessary to change keys periodically (at least every $n^{1/2}$ transmissions) even in the absence of any evidence of eavesdropping, in order to prevent an eavesdropper from intercepting all of the bit positions, a few at a time, over the course of many apparently safe transmissions. The expected number of safe key changes has not been worked out,  but it is not implausible that it is still exponential in the key size.\footnote{\,Decades after these words were written, the basic idea behind this paper was reinvented \emph{independently} by Ivan Damg{\aa}rd, Thomas Pedersen and Louis Salvail, but they worked out the complete analysis of its security, which is missing here. Fittingly, these two papers will appear together in a special issue of \emph{Natural Computing} celebrating 30~years of~BB84.}
\vspace{1ex}

\pagebreak

\section*{Acknowledgements} %TYPO: was "Acknowlegdments" in the original BBB82

We wish to thank Stephen Wiesner for numerous helpful discussions %TYPO: was "disussions" in the original BBB82
of quantum theory, John Denker for drawing our attention to the analogy between choice of basis (e.g.~rectilinear vs.\ diagonal) and a cryptographic key, \mbox{Andrew} Greenberg for pointing out that photons in flight could be used to test a channel for eavesdropping, and Lalit Bahl for advice on error-correcting codes.\footnote{\,We also thank Ilana Frank Mor for typesetting this paper in 2014 from the original 1982 manuscript and for detecting most of the typographical mistakes that have been corrected here.}

%\newpage

\end{document}